\documentclass[conference]{IEEEtran}
\IEEEoverridecommandlockouts
\usepackage{cite}
\usepackage{amsmath,amssymb,amsfonts}
\usepackage{algorithmic}
\usepackage{graphicx}
\usepackage{textcomp}

\usepackage{xcolor}
\def\BibTeX{{\rm B\kern-.05em{\sc i\kern-.025em b}\kern-.08em
    T\kern-.1667em\lower.7ex\hbox{E}\kern-.125emX}}
\begin{document}

\title{Simplified Design Approach for Via Transitions up to 67 GHz \\

}

\author{

\IEEEauthorblockN{ Giorgi Tsintsadze, Reza Vahdani, James L. Drewniak}
\IEEEauthorblockA{\textit{Missouri Science and Technology} \\
\textit{EMC LAB}\\
Rolla, Missouri \\
gt759@mst.edu, rvp7k@mst.edu, drewniak@mst.edu}

\and

\IEEEauthorblockN{Richard Zai}
\IEEEauthorblockA{\textit{Packetmicro} \\
Santa Clara, CA \\
rzai@packetmicro.com}

}

\maketitle

\begin{abstract}
A systematic approach for high-speed via transition design is proposed. The effects of via barrel radius, anti-pad size, and the distance from adjacent stitching (GND) vias on bandwidth are analyzed and characterized. Guidelines for selecting parameter values are provided and validated by correlating 3D full-wave FEM simulation results with actual measurements of the coupon board. When a sufficient number of stitching vias are used, the via structure can be approximated as a coaxial transmission line. The proposed methodology builds on this approximation and also considers high-order modes. With this framework, engineers can easily optimize design parameters while intuitively understanding how geometry affects bandwidth. This approach also allows engineers with limited access to expensive and computationally intensive 3D FEM tools to design high bandwidth vias up to 67 GHz. 
\end{abstract}


\section{Introduction}

A high-bandwidth via is a backbone for the modern multi-layered high-speed printed circuit boards (PCB) \cite{b1}. Due to recent emerging speed of PAM4 224 Gbps, which has Nyquist frequency of 53.123 GHz, design considerations demand that via transition should support signal transmissions up to these frequencies and for material characterization even higher which is typically 67 GHz. Designing a multi gigahertz via is not a trivial matter and has been addressed by number of researchers, for example \cite{b3} considers powerful analytical via modeling up to 40 GHz, however reference vias (GND) are not taken into considerations, while it simplifies analytical model and even though the model correlates to measurement up to 40 GHz, actual transmission bandwidth is substantially lower due to missing reference vias,  at the same time, designing via transition without proper reference (GND) vias is not practical. The goal of this paper is to simplify via geometry to the point where basic transmisison line and waveguide theory can be used to successfully design high-speed via in short time. On top of it, this paper gives engineers a simple but powerful framework to design multi gigahertz via transitions even without access to 3D FEM solvers. 

\section{Coaxial Approximation}
\label{section:coaxial}

\begin{figure}[htbp]
  \centering
  \includegraphics[width=0.5\textwidth]{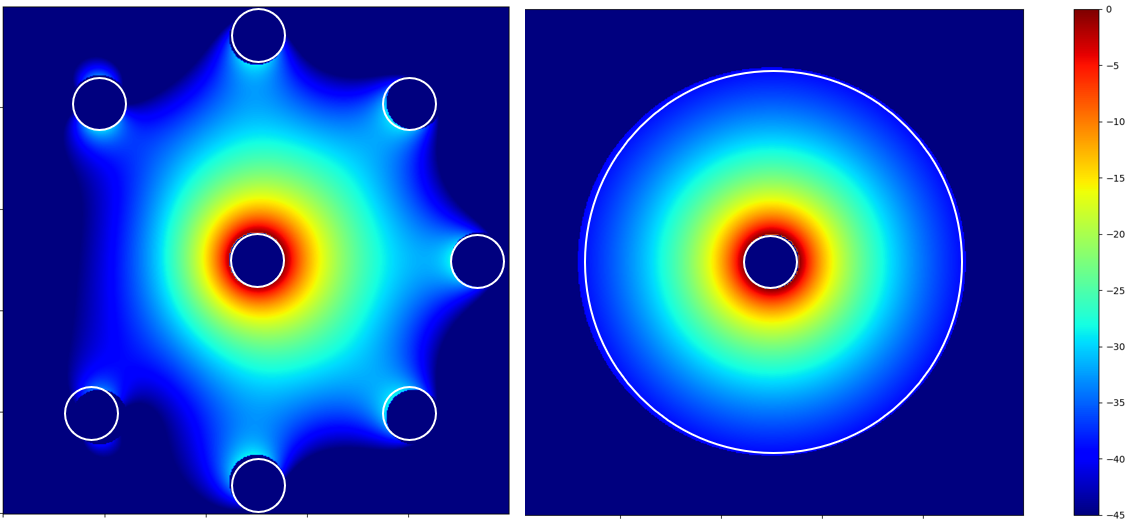}
  \caption{Comparison of Normalized E-field in Log Scale}
  \label{fig:e_field}
\end{figure}

\begin{figure}[tb]
  \centering
  \includegraphics[width=0.3\textwidth]{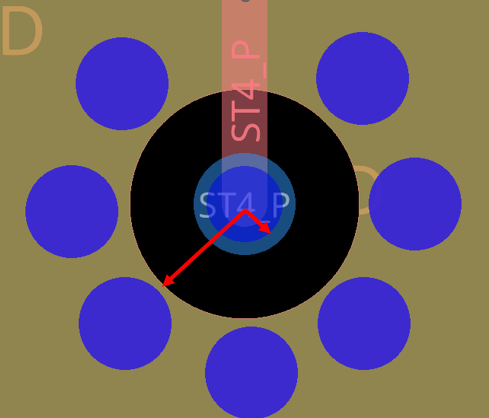}
  \caption{Top 2D view of via structure}
  \label{fig:2d_top}
\end{figure}

\begin{figure}[tb]
  \centering
  \includegraphics[width=0.35\textwidth]{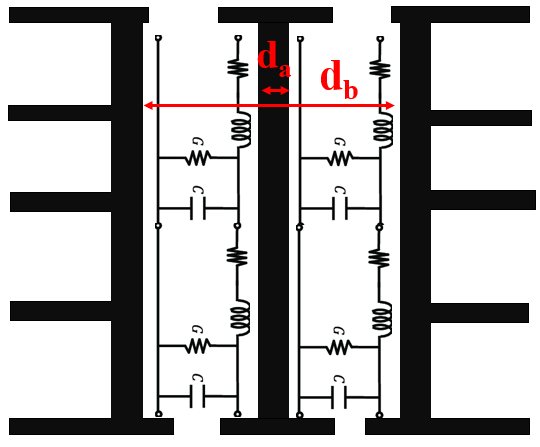}
  \caption{2D Cross section of Via when anti-pad radius is same as distance from stitching via to center via}
  \label{fig:via_coax_init}
\end{figure}

The via structure can be approximated by a coaxial transmission line when sufficient number of stitching vias are present and they are equally spaced from the signal via center, additionally, anti-pad radius is set to be the same as GND via distance from the center via. While not the final geometry, this is a good starting point to understand via transition characteristics in a simplified way. Fig. \ref{fig:2d_top} shows top view of such anti-pad/stitching via placement while Fig. \ref{fig:via_coax_init} shows 2D cross section of the via transition structure. Additionally, Fig. \ref{fig:e_field} shows normalized E-field distribution when 7 stitching vias are used in comparison to ideal coaxial field distribution.

Two main characteristics of a circular coaxial transmission line are it's characteristic impedance and cutoff frequency for high-order modes. From \cite{b2} characteristic impedance is given as:

\begin{equation} \label{eq:coax_impedance}
Z_0 = \sqrt{\frac{\mu}{\epsilon}} \frac { \ln{(\frac{b}{a})} } {2 \pi}
\end{equation}

where a is the inner radius and b is the outer radius of the coaxial line. To achieve a desired $Z_0$, first, inner radius (via barrel radius) can be chosen and the value of outer conductor radius, which in this case is distance from center via to stitching via, can be calculated from (\ref{eq:coax_impedance}). While choosing via barrel radius however, one of the biggest objectives is to make sure the structure is only supporting TEM modes up to the frequency of interest. for coaxial transmission line, cutoff frequency is given as:

\begin{equation} \label{eq:cutoff_frequency}
f_{c} = \frac{ c k_c }{2 \pi \sqrt{\epsilon_r}}
\end{equation}

where c is speed of light in free space, $k_c$ is cutoff wavenumber and approximately can be calculated as:

\begin{equation} \label{eq:cutoff_wavenumber}
k_{c} = \frac{2}{a  + b}
\end{equation}

Eq. (\ref{eq:cutoff_wavenumber}) is an approximate value of cutoff frequency, the more accurate result can be achieved by solving Helmholtz's equation in cylindrical coordinates which has solution in terms of Bessel's functions, however, when enforcing boundary conditions, resulting characteristic equation to obtain $k_c$ must be solved numerically \cite{b2}.

\section{Measurement Results}

\subsection{Design Flow}
\begin{figure}[htbp]
  \centering
  \includegraphics[width=0.35\textwidth]{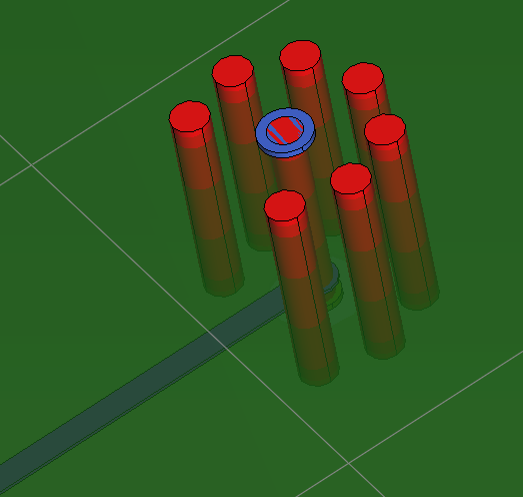}
  \caption{3D view of via transition structure}
  \label{fig:3d_via}
\end{figure}

\begin{figure}[htbp]
  \centering
  \includegraphics[width=0.5\textwidth]{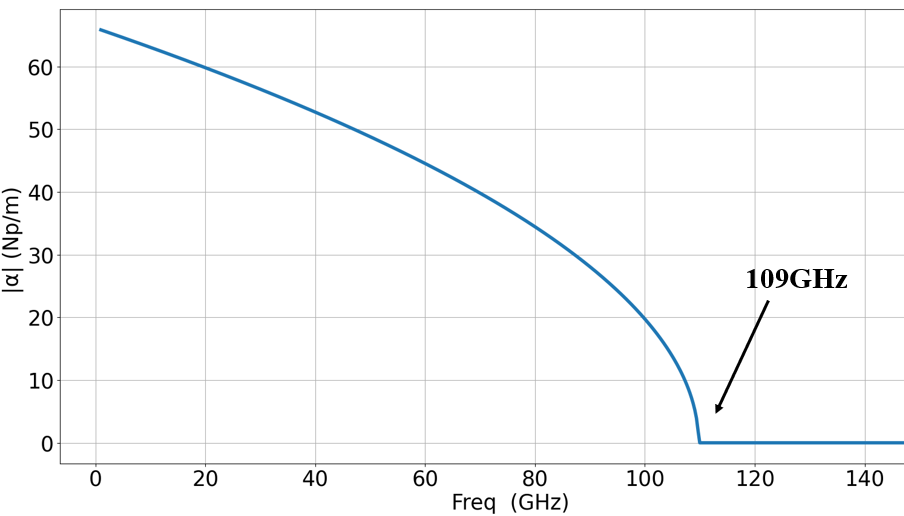}
  \caption{Real part of propagation constant w.r.t frequency for via dimension as via barrel radius of 3.5mils, distance from center via to stitching via edge 15mils.}
  \label{fig:attenuation}
\end{figure}

\begin{figure}[htbp]
  \centering
  \includegraphics[width=0.5\textwidth]{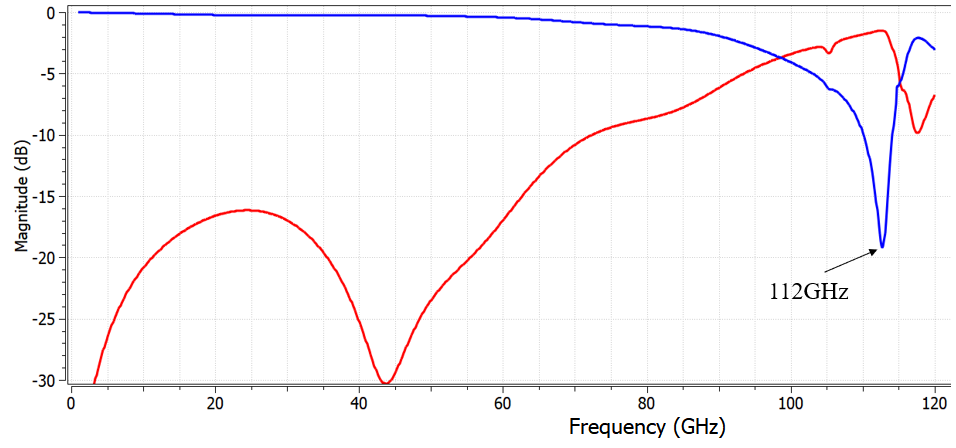}
  \caption{S-parameters of initial via design}
  \label{fig:s_param_initial}
\end{figure}

\begin{figure}[htbp]
  \centering
  \includegraphics[width=0.5\textwidth]{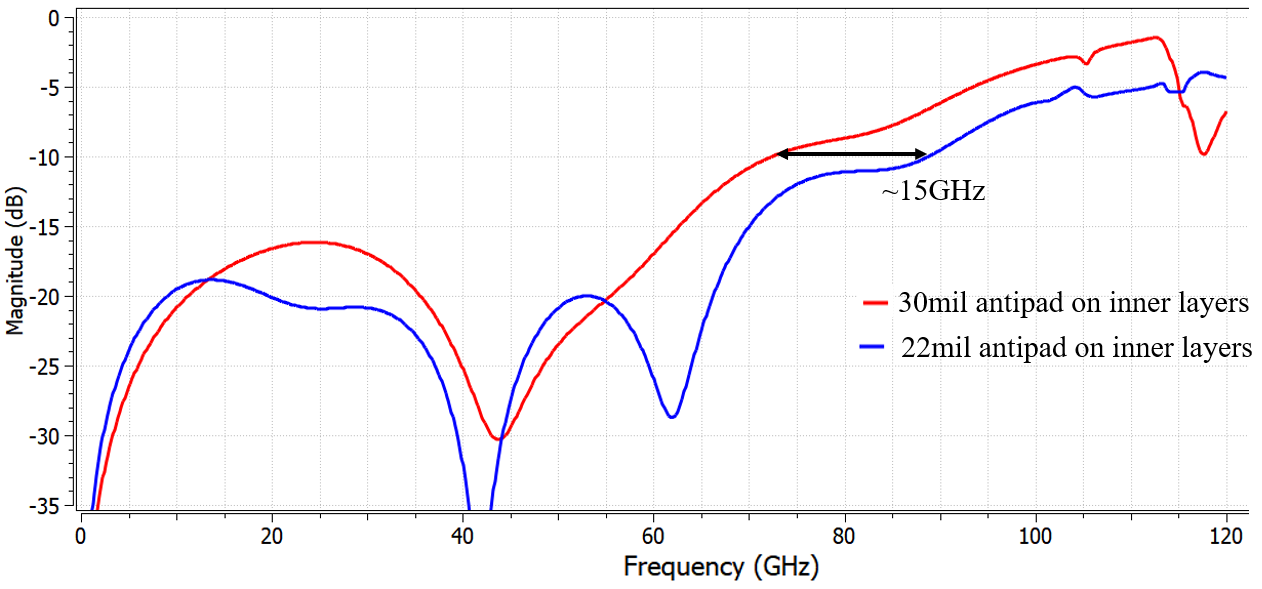}
  \caption{Comparison $S_{11}$ }
  \label{fig:s_param_comparison}
\end{figure}

\begin{figure}[htbp]
  \centering
  \includegraphics[width=0.35\textwidth]{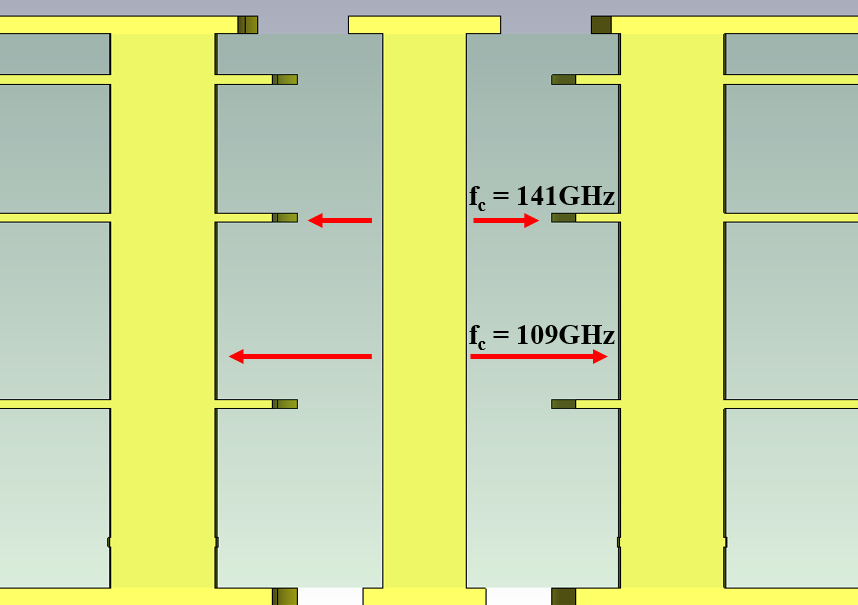}
  \caption{Cross section of updated via structure with 22 mils anti-pad radius on inner layers }
  \label{fig:2d_antipad_modulation}
\end{figure}

In this section, a practical example of via transition design flow is given. Fig. \ref{fig:3d_via} shows geometry of the via that breaks to the stripline. For visualization purposes, reference planes are hidden and only stitching vias, signal via and stirpline is shown while dielectric layers are transparent. In this case, due to manufacturing capabilities and 1.85 mm precision connector dimensions, anti-pad is 30mils, while via barrel diameter is 7mils and $\epsilon_r$ = 3.62 resulting in $Z_0 = 42.5\Omega$ while cutoff frequency $f_c$ according to Eq. (\ref{eq:cutoff_frequency}) is 109 GHz. Fig. \ref{fig:attenuation} shows real part of propagation constant $\gamma$ responsible for attenuation for $TM_{01}$ mode. Fig. \ref{fig:s_param_initial} shows S-parameters of such via structure described above. Notice that there is no transmission at 112 GHz which is very close to predicted value of 109 GHz by approximating via structure with an ideal coaxial transmission line, one other important characteristics is that even though cutoff frequency is 109 GHz, $S_{11}$, takes value of -10 dB at 70 GHz, essentially having much lower effective bandwidth than expected from cutoff frequency. The reason is that even though cutoff frequency is 109 GHz, frequencies that are not supported are not immediately attenuated by the structure, as Fig. \ref{fig:attenuation} shows, attenuation constant is a strong function of frequency. To overcome this issue and extend the effective bandwidth, inner layer antipad size can be reduced so that for inner layer cutoff frequency will be substantially higher resulting in higher attenuation to those frequencies that are not supported but the obvious trade-off is multiple reflections and possible manufacturing capability issues. Fig. \ref{fig:s_param_comparison} shows comparison of $S_{11}$ when inner layer anti-pad radius is changed from 30 mils to 22 mils. A gain of 15 GHz in effective bandwidth, Fig. \ref{fig:2d_antipad_modulation} shows cross section of such via structure with it's associated cutoff frequencies.

\subsection{Effects of Via Barrel Diameter Size}

\begin{figure}[htbp]
  \centering
  \includegraphics[width=0.5\textwidth]{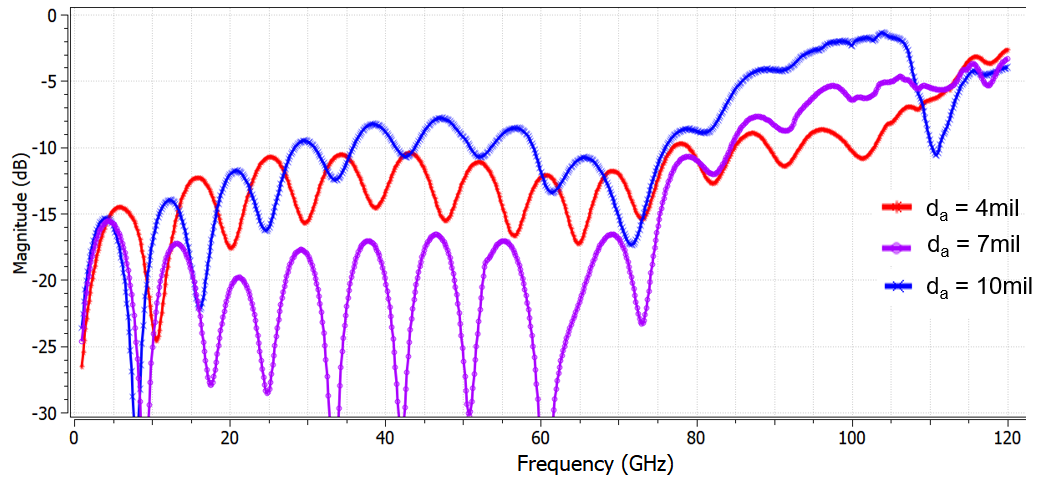}
  \caption{$S_{11}$ of via transition with via barrel diameters of 4 mil, 7 mils and 10 mils corresponding to coaxial transmission line characteristic impedance of 62 $\Omega$, 42.5 $\Omega$ and 36 $\Omega$ respectively}
  \label{fig:s11_different_barrel}
\end{figure}

As mentioned in subsection \textit{A} of section [\ref{section:coaxial}], to maximize via transition bandwidth, it is necessary to limit propagation modes only to TEM. As described in \cite{b4}, magnetic frill currents can excite $TM_{0n}$ modes. Rigorous derivation is presented in \cite{b4}. The high-order modes propagate at different speed and have substantially different propagation constant compared to TEM waves, additionally voltages and currents are not well defined. As a rule of thumb, to increase cutoff frequency of the transition, the structure must be small with respect to wavelength, however, while shrinking via barrel radius indeed increases cutoff frequency, it also changes characteristic impedance of approximate coaxial structure, which on the other hand has negative effects on bandwidth. Fig. \ref{fig:s11_different_barrel} shows comparison of $S_{11}$ of the via structure when diameter of via barrel is 4 mil, 7 mils and 10 mils corresponding to characteristic impedance of 62 $\Omega$, 42.5 $\Omega$ and 36 $\Omega$ respectively. Antipad sizes, stitching via location and number is unchanged and is same as used in section \ref{section:coaxial}. Impedance of the stripline that the via breaks out out to is 42.5 $\Omega$. While figure [\ref{fig:s11_different_barrel}] shows that $S_{11}$ is the best for the via diameter when characteristic impedances of via coaxial approximation is matched to stripline impedance, it is also worthwhile to note that via with 4 mils barrel diameter has superior performance above 85 GHz, the reason is that, while not matched, cutoff frequency is higher and effects from high-order modes are seen later in frequency domain. 

\begin{figure}[htbp]
  \centering
  \includegraphics[width=0.5\textwidth]{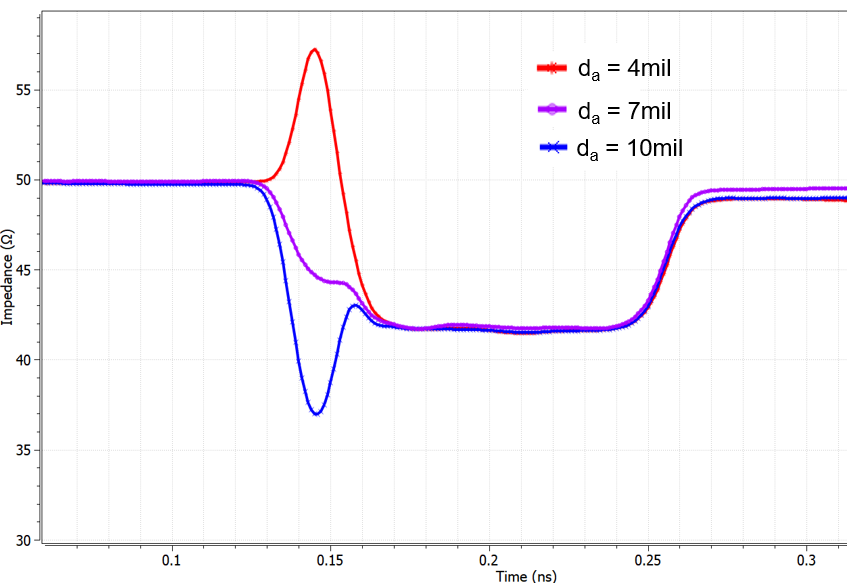}
  \caption{TDR of via transition with via barrel diameters of 4 mil, 7 mils and 10 mils corresponding to coaxial transmission line characteristic impedances of 62 $\Omega$, 42.5 $\Omega$ and 36 $\Omega$ respectively}
  \label{fig:tdr_4_7_10}
\end{figure}

While frequency domain analysis is well suited for via bandwidth characterization, it is also important to consider a time domain analysis, namely time-domain reflection (TDR), to understand discontinuities at the via transition geometry. When designing via transition, engineers' goal is to minimize discontinuities, meaning that transition from precision press-fit coaxial connectors to a stripline with arbitrary characteristic impedance, ideally, should not have inductive peaks or capacitive dips. Fig.  \ref{fig:tdr_4_7_10} shows TDR of via transitions with identical barrel diameters used in Fig. \ref{fig:s11_different_barrel} to show effects of mismatch in time domain. The results are in agreement with guidelines suggested in section \ref{section:coaxial}, transmission line theory can be used to understand via barrel diameter effects on bandwidth and TDR discontinuities, for the value of $d_a$ that has lower characteristic impedance than the stripline, via transition is capacitive hence the dip in the TDR, on the other hand, if $d_a$ is chosen such that characteristic impedance is higher than the stirpline, via transition will look as an inductive discontinuity.

To summarize the subsection, essentially, via barrel diameter should be chosen such that it exerts desired characteristic impedance on the via structure approximated as a coaxial transmission line, at the same time, $TM_{0n}$ cutoff frequency must be kept in mind.

\section{Measurement Results}

\subsection{Test Vehicle Board Overview}

\begin{figure}[htbp]
  \centering
  \includegraphics[width=0.5\textwidth]{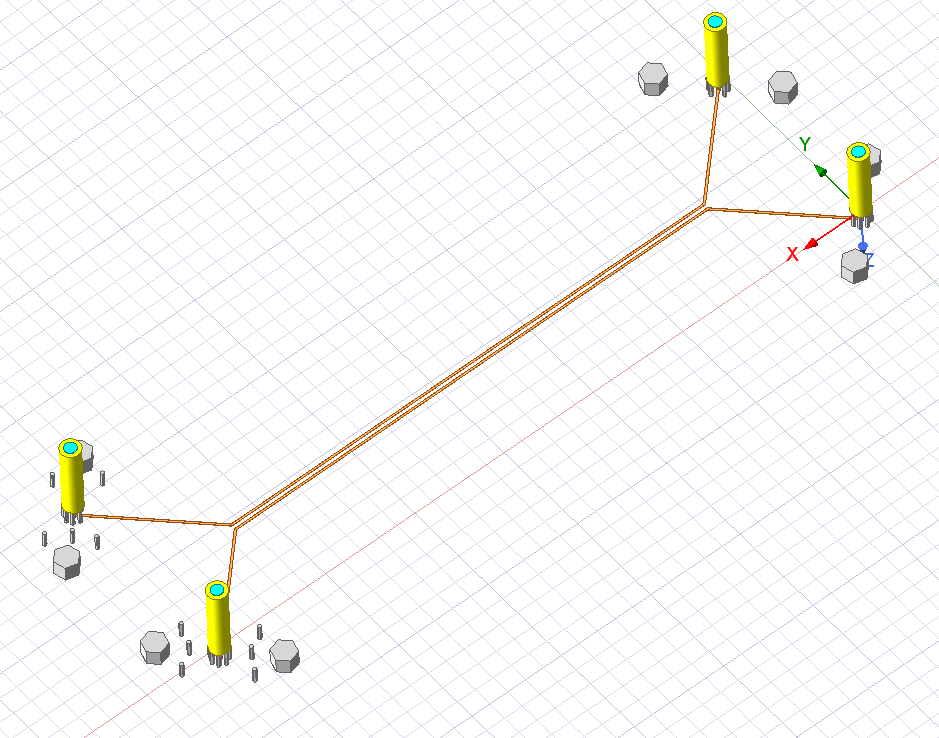}
  \caption{3D model of stripline in 3D FEM solver where dielectric planes and conductor plates are hidden for visualization purposes.}
  \label{fig:ansys3D_model}
\end{figure}

\begin{figure}[htbp]
  \centering
  \includegraphics[width=0.5\textwidth]{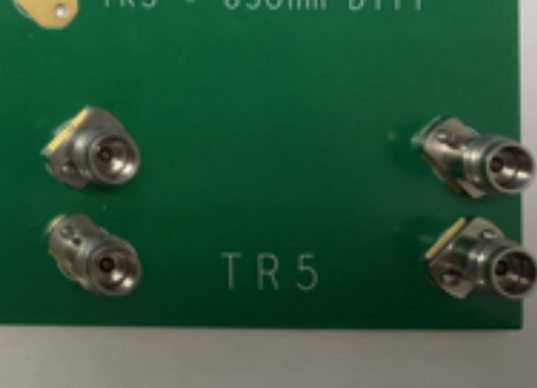}
  \caption{Device under test connected to VNA using press-fit connectors}
  \label{fig:DUT_image}
\end{figure}

\begin{figure}[htbp]
  \centering
  \includegraphics[width=0.5\textwidth]{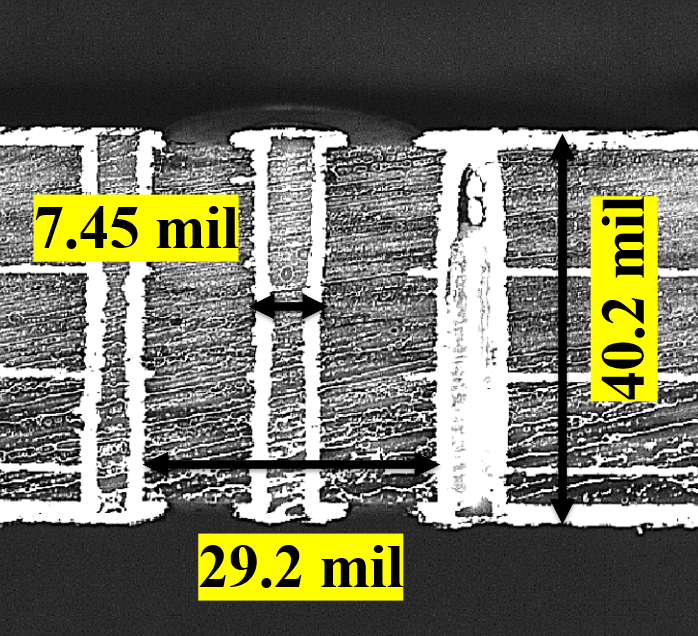}
  \caption{Cross section of via transition of the test vehicle PCB}
  \label{fig:x_section_reza}
\end{figure}

To verify simulation results and general guidelines for designin high-bandwidth via, a test vehicle PCB has been built. the DUT for characterizing via design is 6 layer board where via is used for transitioning from TOP layer to 5th layer as 6th layer acts as a reference plane (GND). Fig. \ref{fig:DUT_image} shows DUT with precision press-fit connectors mounted and connected to VNA while Fig. \ref{fig:ansys3D_model} shows 3D model of vias and traces only by hiding reference planes and dielectrics for visualization purposes. We note that ideal coaxial structures are used in 3D simulation to feed excitation and the length is chosen to match the electrical length of the precision press-fit connector in this case. 
Since one of the goal of this paper is to validate methodology by correlating simulation and measurement, it is necessary to cross-section via geometry and inspect dimensions under the microscope. Fig. \ref{fig:x_section_reza} shows cross section of the via transition and corresponding dimensions. The inner barrel diameter is ~7.2 mils, which is very close to design value of 7 mils. On the other hand, distance between edge of reference via (GND) to signal via center is ~29.45 mils which also matches closely intended value of 30 mils.

\subsection{Simulation and Measurement Correlation}

\begin{figure}[htbp]
  \centering
  \includegraphics[width=0.5\textwidth]{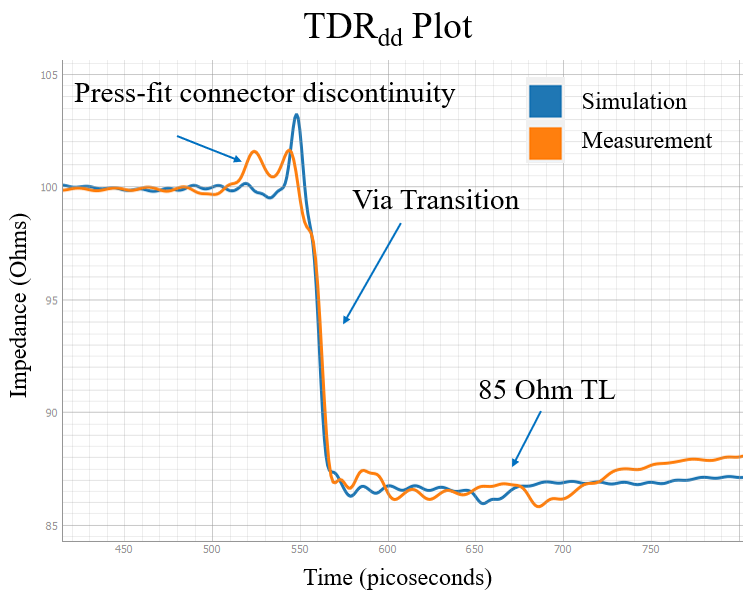}
  \caption{Differential TDR comparison for simulation and measurement.}
  \label{fig:TDRdd_comparison}
\end{figure}

 After validating that cross-section of via matches that of design values, 3D FEM simulation with Ansys HFSS was performed to calculate S-parameters. While S-parameters are powerful to characterize electromagnetic response of a system, directly comparing Sdd11 of the measurement to simulation is not the best way for correlation. The reason is that Sdd11 is a combination of multiple discontinuities and is affected by poor impedance control from manufacturer’s side in some cases, to isolate via transition effects alone in frequency domain is challenging, due to this reason, Sdd11 has been converted to time domain step response also known as TDR. Fig [\ref{fig:TDRdd_comparison}] shows differential TDR comparison of simulation and measurement. In Fig [\ref{fig:TDRdd_comparison}] it is easy to distinguish where via transition happens from press-fit connector (or in case of simulation coaxial feed line) to the 85 Ohm differential stripline. The simulation and measurement have practically identical slope of descent emphasizing the accuracy of the proposed design guideline.

\section{Conclusion}

This paper has described how to approximate via geometry to a transmission line and use simplified methodology to design high-bandwidth via transition even without the 3D FEM tools or complicated analytical modelling. This framework helps engineers understand mechanism of reflection in the via transition whether it is due to characteristic TEM impedance mismatch or geometry supporting high-order modes. The theory has been validated though measurement and simulation correlation where it showed excellent correlation. As for the limitations, this paper only considered homogeneous stackup scenario where dielectric constant is same for each layer, however some hybrid stack-ups utilize different Dk materials that will need slightly different treatment but this is reserved for future research. 

\section*{Acknowledgment}

This work was supported in part by the National Science
 Foundation (NSF) under Grant IIP-1916535

\vspace{12pt}
\color{red}

\end{document}